\documentclass[prb,twocolumn,superscriptaddress,showpacs,amsmath,amssymb]{revtex4}
\usepackage{amssymb}
\usepackage{amsmath}
\usepackage{graphicx}
\usepackage{bm}
\begin{document}
\title{Magnetic-field-induced transition from metastable spin glass to possible antiferromagnetic-ferromagnetic phase separation in $Cd_{0.5}Cu_{0.5}Cr_2O_4$}

\author{Li-qin Yan}
\email{lqyan@aphy.iphy.ac.cn}
\author{Wen Yin}
\affiliation{National Laboratory for Condensed Matter Physics,Institute of Physics, Chinese Academy of Sciences, Beijing 100080,China}
\author{Ferran Maci\'a}
\affiliation{Departament de F¨ªsica Fonamental, Facultat de F¨ªsica, Universitat de Barcelona, Avda. Diagonal 647, Planta 4, Edifici nou, 08028 Barcelona, Spain}
\author{Jun-rong Zhang}
\author{Lun-hua He}
\author{Fang-wei Wang}
\affiliation{National Laboratory for Condensed Matter Physics,Institute of Physics, Chinese Academy of Sciences, Beijing 100080,China}

\date{\today}

\begin{abstract}
Using ac susceptibility, dc magnetization and heat capacity measurements, we have investigated the magnetic properties of $Cd_{0.5}Cu_{0.5}Cr_2O_4$. $Cd_{0.5}Cu_{0.5}Cr_2O_4$ has an extraordinary magnetic phase including a metastable spin-glass(SG) phase at zero field, a possible phase separation scenario of AFM/FM above $\sim 0.5T$ field, and at intermediate fields, an apparent pseudo reentrant spin-glass (RSG) plateau is observed. These phenomena are closely correlated with the pinning effect of the $Cu^{2+}$ sublattice on the frustrated lattice.
\end{abstract}

\pacs{75.50.Lk, 75.30.Kz, 75.50.Ee}
\keywords{Spin glass, Magnetism, $Cd_{1-x}Cu_xCr_2O_4$}
\maketitle

\section{introduction}
The notion of a geometrically frustrated antiferromagnet\cite{1}  has attracted considerable interest over the past more than one decade. In its simplest form, a lattice geometry results in frustration of the antiferromagnetic (AFM) exchange interaction. Such materials are characterized by the absence of long range order at temperatures well below the Curie-Weiss temperature ($\Theta_{CW}$), and have very unusual low temperature properties\cite{2,3,4,5,6,7,8,9,10,12}. An interesting feature of the system is the release of magnetic frustration. Normally, the magnetic frustration can be eliminated or removed by spin-Peierls transition at $T_N$ arising from the spin-lattice coupling or an applied high magnetic field at low temperature\cite{2,3,4} . Theoretically, the vast degeneracy of their classical ground states makes them be highly susceptible even to small perturbations. A separate phase transition into an AFM state is expected\cite{5}.Experiments have shown that by means of replacing the nonmagnetic-site partially by magnetic ions, a ¡°local preferential direction¡± will be imposed in frustrated lattice, removing somehow the strong intrinsic magnetic frustration and presenting a geometric spin-glass(SG) state\cite{6,7} . Note that the presence of the disorder coming from either the site disorder or the competing interaction between AFM and ferromagnetism (FM), often generates a conventional SG state. In systems which possess disorder and highly geometric frustration often display some unconventional SG behavior, usually named as ¡°geometrical SG¡±. If so, then a subtle balance between them, i.e., disorder and geometric frustration, will bring about an unusual SG state. The views on the nature of the freezing of this SG, whether it is a phase transition or a non equilibrium phenomenon, are still controversial. And it is unclear how it will evolve under a small perturbation provided that it is a nonequilibrium SG. Understanding the nature of such a SG state is regarded as an important physics issue from theoretical and experimental points of view\cite{8,9}. 

$CdCr_2O_4$ is a classical frustrated antiferromagnet, in which the magnetic Cr ions are on three dimensional corner-sharing tetrahedral sublattices, which results in geometrical frustration of the AFM nearest neighbor exchange interactions ($T_N$, 7.8 K)\cite{10,11} . It undergoes a first-order three-dimensional spin-Peierls transition at $T_N$ from a cubic paramagnetic to a tetragonal N¨¦el state\cite{10}; Low-temperature neutron powder diffraction showed the presence of spiral AFM spin structure\cite{11}  and muon-spin-relaxation measurements (mSR) indicated that substantial magnetic frustration still remains at the millikelvin temperatures\cite{12}. In this paper, we report on the SG state of a high-quality polycrystalline $Cd_{0.5}Cu_{0.5}Cr_2O_4$ with magnetic ions $Cu^{2+}$ half doped revealed by ac susceptibility, dc magnetization and heat capacity measurements. We demonstrate for the first time that the ground-state is well characterized in a metastable SG state at zero field and a possible magnetic phase separation of AFM/FM under a magnetic field as the ground-state degeneracy is broken.
\section{experiment}
The preparation, crystal structure and the primary magnetic property of $Cd_{0.5}Cu_{0.5}Cr_2O_4$ in present study have been well described and characterized by previous experimental measurements\cite{13}. All magnetization and heat capacity measurements were performed using a Physical Property Measurement System (Quantum Design). Data were collected upon warming after cooling the samples at zero field. Specific-heat measurements were done in a thermal relaxation method.
\section{results}
\subsection{Structure and fundamental magnetic property}
The crystal structure of $Cd_{0.5}Cu_{0.5}Cr_2O_4$ has a cubic spinel space group $Fd3m$, which consists of two basic units, $Cd/CuO_4$ tetrahedron and $CrO_6$ octahedron. There are at least two main kinds of magnetic interactions in $Cd_{0.5}Cu_{0.5}Cr_2O_4$, the nearest-neighbor Cr-Cr interaction and the next nearest-neighbor Cu-Cr interaction, according with reentrant SG (RSG) character where only the first and second nearest-neighbor interactions are important in RSG materials\cite{14}. The Cu-Cr and Cd-Cr bonds are not uniformly distributed due to A-site random distribution. Clearly, upon substitution of Cd by magnetic Cu ions, the competition between the Cu-Cr and Cr-Cr interactions causes spin arrangements transfer from AFM to ferrimagnetic and the geometrical frustration turns to be suppressed\cite{13}. Thus the $Cu^{2+}$ magnetic sublattice is extracted with ferromagnetic coupling, similar to that of tetrahedral spinel $CuCr_2O_4$, where the magnetic moment of the Cu sublattice is ferromagnetic coupling, antiparallel to the resultant Cr sublattice one\cite{15}. Previous ZFC magnetic study at 0.1T\cite{13}indicated that $Cd_{0.5}Cu_{0.5}Cr_2O_4$ experiences a paramagnetism(PM)-ferrimagnetism(FI)-SG (RSG) like transition ($T_c$, 20K; $T_g$, 8K). At the same time, its index of frustration ($f ={|\Theta_{cw}/T_N|}$) varies from $~10$ to $~1$, which might be attributed to the interaction between $Cu^{2+}$ sublattice and $Cr^{3+}$ sublattice, namely, the ¡°pinning effect¡± of sublattice $Cu^{2+}$ on magnetic frustration (or spiral AFM configuration) excites a ¡°local preferential direction¡±, similar to Mn-rich $YMnO_3$\cite{6} and $SrCr_8Ga_{4-2x}Fe_xO_{19}$ series\cite{7}.
\begin{figure}
\includegraphics[width=0.4\textwidth]{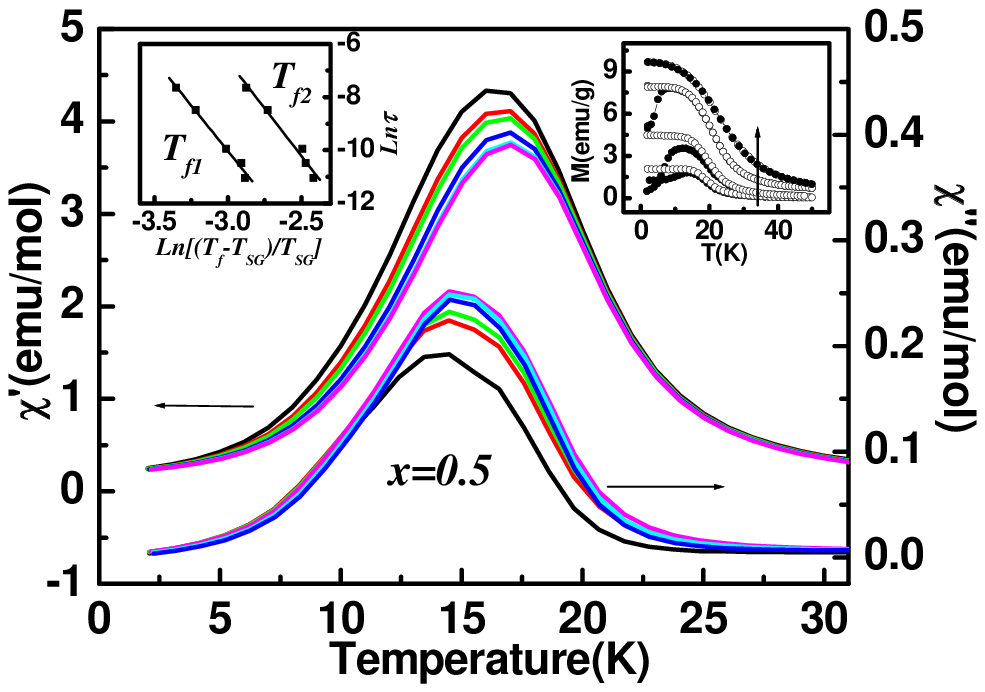}
\caption{(Color online) $\chi^{'}(\omega)$ and $\chi^"(\omega)$  vs T for $\omega /2\pi$ = 33,333,777,3333,7777,9999 Hz[top to bottom for $\chi^{'}$  and bottom to top for $\chi^"$]. The left inset displays the measured freezing temperatures $T_{f1}(\omega,T)$, $T_{f2}(\omega,T)$ and the best fitted line by Eq.(1) for $\chi^{'}$ and $\chi^"$. The right inset shows the temperature dependence of ZFC (closed circles) and FC (open circles) magnetization for $Cd_{0.5}Cu_{0.5}Cr_2O_4$ compound under various magnetic fields of 0.01, 0.05, 0.1, 0.5 T. The arrow indicates the direction of increasing field.}
\end{figure}
\subsection{Metastable spin glass}
Figure 1 shows the temperature dependence of ac susceptibility data in a frequency range of $33\leqslant\omega/2\pi\leqslant9999Hz$ under ac field of 1Oe for  $Cd_{0.5}Cu_{0.5}Cr_2O_4$. The curves display a maximum at a temperature $T_f$, which shifts with (increasing) frequency upwards for $\chi'(\omega,T)$ and downwards for $\chi^"(\omega,T)$. This is a distinct feature of a SG state\cite{16}. The value of the frequency sensitivity of $T_f(\omega)$,$\Delta T_f(\omega)/[T_f(\omega)\Delta log_{10}\omega]$, has been a criterion for the presence of a canonical SG from SG like\cite{17}. It is about 0.013 for $\chi^{'}(\omega,T)$, lower than those reported for other typical insulating SG systems, indicating an unconventional SG transition. However, it is about 0.025 for $\chi^"(\omega,T)$, close to that of conventional SG.\cite{18}. We know the out-of-phase $\chi^"$,  is the magnetic energy loss, sometimes reflecting certain information that is drown or not obvious in the in-phase $\chi^{'}(\omega,T)$. The different extracted parameters for $\chi^{'}(\omega,T)$ and $\chi^"(\omega,T)$ appear to reflect that this unusual SG in $Cd_{0.5}Cu_{0.5}Cr_2O_4$ is made by more than one component. The divergence of the maximum relaxation time $\tau_{max}$, occurring at the peak temperature, can be investigated by using conventional critical slowing down:
\begin{equation}
\frac{\tau}{\tau_0}=\xi^{-z\nu}=\bigglb(\frac{T_f(\omega)-T_{SG}}{T_{SG}}\biggrb)^{-z\nu}
\end{equation}
here, $T_{SG}$ is the SG phase transition temperature and $T_{f1}$ and $T_{f2}$, defined as the maxima of the in-phase and out-of-phase ac susceptibility, respectively, are the frequency-dependent freezing temperatures at which the maximum relaxation time ¦Ó of the system corresponds to the measured frequency. The left inset of Fig.1 presents a best fit to the data. When $\tau_0$ is $10^{-13}$ s typically taken in the SG system, $T_{SG}$ and $z\nu$ for $\chi^{'}(\omega,T)$ are 15.90 K and $z\nu$ = 6.08, respectively, whereas, for $\chi^"(\omega,T)$, a good scaling yields $T_{SG}$ = 13.15 K and $z\nu$ = 7.80. z and $\nu$ are critical dynamics exponents. Although the $T_{SG}$ and $z\nu$ values of $\chi^{'}(\omega,T)$ and $\chi^"(\omega,T)$ are slightly different, their magnitudes of $z\mu$ are within the conventional SG phase transition\cite{19}. 

It is reminiscent that the appearance of ¡°RSG¡±-like plateau in M-T curve of our previous study\cite{13} may be induced by an applied dc magnetic field since no RSG plateau is observed in the in-phase $\chi^{'}(\omega,T)$ as well as out-of-phase $\chi^"(\omega,T)$. This assumption is further confirmed by the temperature dependence of the zero-field-cooled (ZFC) and field-cooled (FC) magnetizations curves from 2 to 300 K in various fields of 0.01-0.5 T (see the right inset of Fig. 1). The cusp of ZFC magnetization at 0.01 T coincides well with the cusp seen in the lowest frequency ac susceptibility measurement (33Hz). The FC and ZFC magnetization curves separate at around 17 K and 0.01 T, then the cusp temperature decreases monotonically with increasing applied field, together with a formation of a pseudo RSG plateau (see the ZFC curves at 0.05 and 0.1 T). Clearly, applying magnetic field in the SG state produces a reduction in this SG phase and an increase of magnetic ordered phase. This transition can be considered as a metastable SG to magnetic ordering transition up to 0.5 T and the superposition of ZFC and FC is observed.
\subsection{Magnetic-field-induced possible AFM/FM phase separation behavior}
\subsubsection{Magnetic field dependence of ac susceptibility}
Figure 2 shows the ac susceptibility $\chi^{'}(\omega,T)$ and $\chi^"(\omega,T)$ for 333Hz, in different superposed dc fields and ac field of 1Oe. Both $\chi^{'}(\omega,T)$ and $\chi^"(\omega,T)$ are suppressed drastically at the dc fields. An applied field rounds the peak off and broadens it to a plateau state, then enters it into a double peak structure for $\chi^{'}(\omega,T)$ (see the left inset of Fig. 2). Obviously, applying a dc field, the SG phase is suppressed and ordered magnetic clusters are induced. Whereas $\chi^"(\omega,T)$ is different from the double peak structure of $\chi^{'}(\omega,T)$, the peaks are smeared out in amplitude and shift downwards (see the right inset of Fig.2), showing a characteristic feature of a conventional SG\cite{20}. With increasing magnetic field, the two maxima in $\chi^{'}(\omega,T)$ shift toward opposite directions, implying a possible magnetic phase separation system. Normally, the transition on the high temperature side is ascribed to a formation of field-induced grown size FM clusters embedded in a SG matrix\cite{21} while the transition on the low temperature side is a conventional SG transition. The formation of FM clusters in a SG matrix should arise from the field-induced stepwise connection of small short-range-ordered clusters along the local preferential direction of magnetic ordering. Once it is a true SG transition on the low temperature side, the dc field dependence of the freezing temperature $T_f(H)$ should be scaled by the equation
\begin{equation}
{T_f(H)}=1-bH^\delta
\end{equation}
with $\delta = 2/3$ via the mean-field theory prediction\cite{19,20}. Figure 2(b) plots the experimental values of $T_{f1}(H)$ for $\chi^{'}(\omega,T)$, $T_{f2}(H)$ for $\chi^"(\omega,T)$, and the fitted curves to Eq.(2).The fitted values of the exponent $\delta$ are 0.186 for $\chi^{'}(\omega,T)$ and 0.023 for $\chi^"(\omega,T)$, which is far lower than the typical SG value 2/3. It thus turns out that the phase transition on low temperature side is not a conventional SG transition. It is speculated that this transition should be dominated by the geometrical SG, which is closely related to AFM transition in the frustrated lattice.
\begin{figure}
\includegraphics[width=0.40\textwidth]{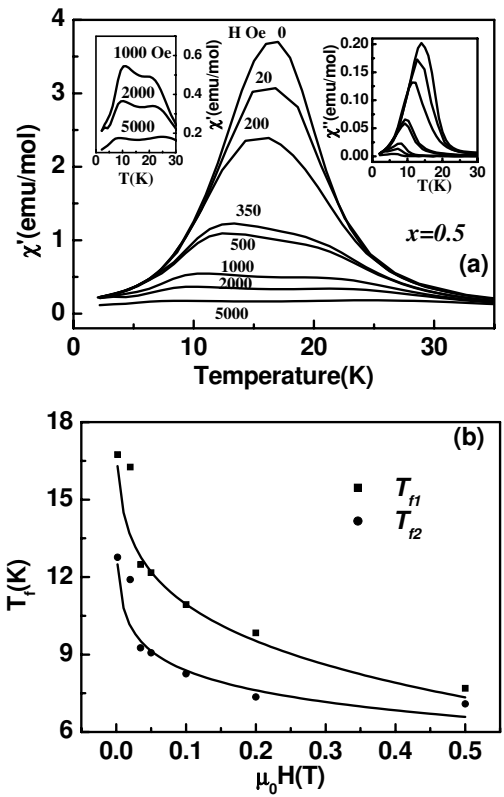}
\label{fig2}
\caption{(a) Temperature dependence of in$-$phase ac susceptibility measured at a frequency of 333 Hz under different applied dc fields. The right inset presents the corresponding out$-$of$-$phase ac susceptibility. The left inset shows a magnification of the in$-$phase ac susceptibility for dc fields $\mu_{0}H$ = 0.1, 0.2, 0.5 T. (b) The experimental $T_{f}(H)$ values and the fitted data to Eq.(2).}
\end{figure}
\subsubsection{Magnetization curve at 2K}
According to the above ZFC and FC measurements, the superposition of ZFC and FC curves takes place at $\mu_0H \geqslant 0.5T$, implying that a completed SG-magnetic ordering transition can be achieved. If dc field leads to a complete transition of SG-ferro(-ferri)magnetic ordering, the magnetic moment should be saturated in high field. However, it is not the case. The inset of Fig. 3(a) presents the original magnetization curve at 2 K. It is obviously hard to saturate and the full saturation state is not achieved even under a field of 13 T, indicating an existence of intrinsic AFM order. The absences of saturation at 13 T and S-type feature can be correlated with the AFM/FM phase separation scenario. The magnetic contribution from ferro(-ferri)magnetic and AFM parts is fitted by a linear least square method, which leads to $M_{ferro} = 0.85 \mu_B/f.u.$ derived from the ferro(-ferri)magnetic lattice part. This absolute value of $M_{ferro}$ is larger than the expected saturation moment $\mu_s \sim 0.5 \mu_B/f.u.$ for the parallel $Cu^{2+}$ ratio at 13 T and 2 K, implying an additive magnetic contribution $\sim 0.35 \mu_B/f.u.$ parallel to $Cu^{2+}$ magnetic moment. This additive magnetic moment should originate from the ¡°local preferential direction¡± in the Cr tetrahedron among the field-induced FM clusters. We have hereby concluded that the magnetic field induces the transition of metastable SG into not only FM clusters but also AFM phase. This transition exhibits a possible phase separation of AFM/FM corresponding to the transition at the ``double peak structures'' in the ac susceptibility measurement under dc fields.
\subsubsection{Magnetic entropy changes from isothermal magnetization measurements}
In order to get an insight into this scenario we carried out the measurements of the isothermal magnetization curves in the temperature range of 5$-$50 K and magnetic fields up to 5.0 T (see Fig. 3(a)). The temperature step of 3 K was chosen. The behavior of the isotherms differs from the typical SG and ferromagnetic behaviors. As the temperature is decreased the $M$ vs $H$ curve bends more, but neither sign of saturation nor S-type behavior is present. Fig. 3(b) shows the Arrot plots obtained from the magnetization isotherms. A considerable curvature above 17 K is observed as a sign of disordered system\cite{22,23}. There is a positive small intercept at 5-14 K and low fields, suggesting that the system exhibits rather a weak magnetization than a conventional SG. The data taken at high fields can be fitted by straight lines. Extrapolating these lines at temperatures between 5 and 14 K, a positive intercept with the $M^2$ axis is reached, indicating a field-induced magnetic order alignment. 
\begin{figure}
\includegraphics[width=0.35\textwidth]{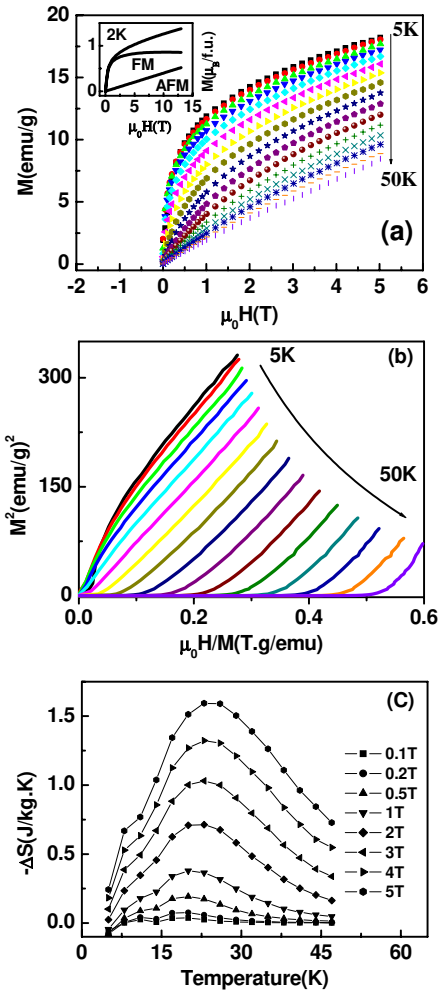}
\label{fig3}
\caption{(Color online) (a) Magnetic field dependence of the magnetization of $Cd_{0.5}Cu_{0.5}Cr_2O_4$ at various temperatures of 5$-$50 K. The inset presents the magnetization curve at 2 K and AFM/FM moment contributions fitted by a linear least square method. (b) The corresponding Arrott plot. (c) Temperature dependence of the entropy changes for the magnetic field difference from 0 to 0.1, 0.2, 0.5, 1, 2, 3, 4, 5 T. The lines are only guides for eyes.}
\end{figure}
Magnetic entropy change versus temperature is shown in Fig. 3(c). Two peaks can be observed at $\sim 11$ and $\sim 21 K$ at $\mu_0H = 0.1$ T, respectively. The kink on the low temperature side shifts from around 11 K to 8 K with the magnetic field, suggesting an ordered AFM transition. Furthermore, the narrow width of $|\Delta S_{M}|$ around $T_{AFM}$ is one of the features of first order phase transitions. On the other hand, the $|\Delta S_{M}|$ peak on the high temperature range is broadened and high fields shifts the Curie temperature several degrees from around 21 K to 24 K, implying now a continuous second ordering FM transition, in agreement with the above result from ac susceptibility under dc field. Consequently, we conclude that the low temperature transition ($\sim 11 K$) is the metastable SG-AFM ordering while the high temperature one ($\sim 21 K)$ is the metastable SG-FM clusters.
\subsubsection{Heat capacity measurement}
To further confirm the scenario discussed above, the measurements of the heat capacity have been performed under the fields of $\mu_0H = 0$ T, and 0.5 T (see Fig. 4)\cite{24}. One advantage of this insulating system is the lack of an electronic contribution to the specific heat.The dependence of $C$ on temperature nearly accords with the $C_m\sim T^2$ law at zero field and low temperatures (the bottom inset of Fig.4), which is similar to geometrically frustrated SG systems, such as the spinel lattice\cite{25} and kagome lattice\cite{8,9}. Furthermore, a character of conventional SG state is observed through a broad maximum of $C/T$ at about 25 K instead of an anomaly at the freezing temperature 17 K at zero field. These two features indicate an unusual metastable SG state in $Cd_{0.5}Cu_{0.5}Cr_2O_4$. No distinct two peaks are observed in the specific heat curve under a field of 0.5 T. Only a knee development is found at $\sim 25$ K. However, the magnetic entropy change $\Delta S_{heat}$ at $\mu_0H = 0.5$ T calculated from the heat capacity data (see the top inset of Fig.4) exhibits the corresponding ¡°double peak structure¡± to $\Delta S_{mag}$, further confirming the possible landscape of AFM/FM phase separation under an external magnetic field.
\begin{figure}
\includegraphics[width=0.40\textwidth]{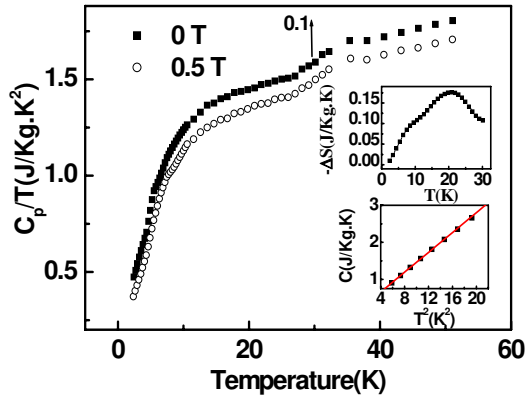}
\label{fig4}
\caption{(Color online) Temperature dependence of the heat capacity of $Cd_{0.5}Cu_{0.5}Cr_2O_4$ measured under the fields of $\mu _{0}H$ = 0 T, and 0.5 T. For clarify, the data for $\mu _{0}H$ = 0 is shifted up $0.1 J/Kg.K^2$. The top inset plots the entropy change from heat capacity measurements with the magnetic field changes from 0 to 0.5 T. The bottom inset depicts the temperature dependence of $C$ deviated from the $C_{m}\sim T^{2}$ law in zero field.}
\end{figure}
\section{DISCUSSION}
Combining our results with the introduction mentioned before, it is already clear that for A-site $Cu^{2+}$ half doped $CdCr_2O_4$, a metastable SG at zero dc field is emergent. The formation of the metastable SG ground state in $Cd_{0.5}Cu_{0.5}Cr_{2}O_{4}$ is closely pertinent to the magnetic interaction between Cu sublattice and Cr sublattice. Half number of Cd atoms substituted by Cu introduces a magnetic moment that increases not only the magnetic interaction between Cu sublattice and Cr sublattice but also the probability of the local $CrO_6$ octahedron distortion. The large ionic radii difference between $Cu^{2+}$ ions($0.72\mathring{A}$) and $Cd^{2+}$ ions ($0.97\mathring{A}$) results in $AO_4$ tetrahedron distortion or a local crystal distortion, which favors the off$-$center of $Cr^{3+}$ ions via oxygen atom displacement. Thus the geometrical frustration tends to release due to the variation of local Cr$-$Cr bond length. This is a benefit for the pinning effect of Cu sublattice on Cr sublattice. The Cu$-$Cr interaction would impose a ``preferential direction'' for the orientation of the spins and therefore, two SG components, namely, conventional SG, which arises from site disorder or magnetic interactions competition, and geometrical SG, which is from the substantial geometrical magnetic frustration, would appear. A subtle balance between them would produce a metastable SG state. 

Applying a magnetic field this balance is destroyed and a transition from metastable SG to possible AFM/FM phase separation is induced. This phenomenon is also closely correlated with the pinning effect of the $Cu^{2+}$ sublattice on a frustrated lattice. The magnetic frustration arises from the $Cr^{3+}$ ions, arranged in a corner$-$sharing tetrahedral lattice (pyrochlore lattice) while the $Cu^{2+}$ sublattice, interacting ferromagnetically, would pin its surrounding magnetic frustration to a preferential direction somehow, exhibiting a metastable SG state. Under a magnetic field the pinning effect will be enhanced and the disorder will become weaker, i.e., the balance between the geometrical SG and conventional SG will be broken, exhibiting AFM/FM phase separation.
\section{CONCLUSIONS}
As a whole, our experimental results provide a magnetic landscape of $Cu^{2+}$ ions half doped $CdCr_2O_4$. $Cu^{2+}$intermediate substitution for $Cd^{2+}$ ions suppresses the magnetic frustration by imposing a ¡°local preferential direction¡± in Cr tetrahedron. The subtle balance state between the pinning interaction and substantial frustration manifests a metastable SG behavior. An applied magnetic field induces a transition from this SG phase to a possible AFM/FM phase separation state. This newly $Cu^{2+}$ doped $CdCr_2O_4$ material will enable the in-depth study on the rich physical properties of magnetic frustration compounds.
\acknowledgments{
One of the authors thanks Professor D. W. Wu, T. Y. Zhao, W. S. Zhan, and H. W. Zhang for their helpful discussions. This work was supported by the National Natural Science Foundation of China (Grant No 10505029).}
\bibliography{yan}

\begin{thebibliography}{25}
\expandafter\ifx\csname natexlab\endcsname\relax\def\natexlab#1{#1}\fi
\expandafter\ifx\csname bibnamefont\endcsname\relax
  \def\bibnamefont#1{#1}\fi
\expandafter\ifx\csname bibfnamefont\endcsname\relax
  \def\bibfnamefont#1{#1}\fi
\expandafter\ifx\csname citenamefont\endcsname\relax
  \def\citenamefont#1{#1}\fi
\expandafter\ifx\csname url\endcsname\relax
  \def\url#1{\texttt{#1}}\fi
\expandafter\ifx\csname urlprefix\endcsname\relax\def\urlprefix{URL }\fi
\providecommand{\bibinfo}[2]{#2}
\providecommand{\eprint}[2][]{\url{#2}}

\bibitem[{\citenamefont{Ramirez}(1994)}]{1}
\bibinfo{author}{\bibfnamefont{A.~P.} \bibnamefont{Ramirez}},
  \bibinfo{journal}{Annu.Rev.Mater.Sci.} \textbf{\bibinfo{volume}{24}},
  \bibinfo{pages}{453} (\bibinfo{year}{1994}).

\bibitem[{\citenamefont{Lee et~al.}(2000)\citenamefont{Lee, Broholm, Kim, II,
  and Cheong}}]{2}
\bibinfo{author}{\bibfnamefont{S.-H.} \bibnamefont{Lee}},
  \bibinfo{author}{\bibfnamefont{C.}~\bibnamefont{Broholm}},
  \bibinfo{author}{\bibfnamefont{T.~H.} \bibnamefont{Kim}},
  \bibinfo{author}{\bibfnamefont{W.~R.} \bibnamefont{II}}, \bibnamefont{and}
  \bibinfo{author}{\bibfnamefont{S.-W.} \bibnamefont{Cheong}},
  \bibinfo{journal}{Phys. Rev. Lett.} \textbf{\bibinfo{volume}{84}},
  \bibinfo{pages}{3718} (\bibinfo{year}{2000}).

\bibitem[{\citenamefont{Penc et~al.}(2004)\citenamefont{Penc, Shannon, and
  Shiba}}]{3}
\bibinfo{author}{\bibfnamefont{K.}~\bibnamefont{Penc}},
  \bibinfo{author}{\bibfnamefont{N.}~\bibnamefont{Shannon}}, \bibnamefont{and}
  \bibinfo{author}{\bibfnamefont{H.}~\bibnamefont{Shiba}},
  \bibinfo{journal}{Phys. Rev. Lett.} \textbf{\bibinfo{volume}{93}},
  \bibinfo{pages}{197203} (\bibinfo{year}{2004}).

\bibitem[{\citenamefont{Ueda et~al.}(2005)\citenamefont{Ueda, Katori, Mitamura,
  and Goto}}]{4}
\bibinfo{author}{\bibfnamefont{H.}~\bibnamefont{Ueda}},
  \bibinfo{author}{\bibfnamefont{H.~A.} \bibnamefont{Katori}},
  \bibinfo{author}{\bibfnamefont{H.}~\bibnamefont{Mitamura}}, \bibnamefont{and}
  \bibinfo{author}{\bibfnamefont{T.}~\bibnamefont{Goto}},
  \bibinfo{journal}{Phys. Rev. Lett.} \textbf{\bibinfo{volume}{94}},
  \bibinfo{pages}{047202} (\bibinfo{year}{2005}).

\bibitem[{\citenamefont{Tchernyshyov et~al.}(2002)\citenamefont{Tchernyshyov,
  Moessner, and Sondhi}}]{5}
\bibinfo{author}{\bibfnamefont{O.}~\bibnamefont{Tchernyshyov}},
  \bibinfo{author}{\bibfnamefont{R.}~\bibnamefont{Moessner}}, \bibnamefont{and}
  \bibinfo{author}{\bibfnamefont{S.~L.} \bibnamefont{Sondhi}},
  \bibinfo{journal}{Phys. Rev. Lett.} \textbf{\bibinfo{volume}{88}},
  \bibinfo{pages}{067203} (\bibinfo{year}{2002}).

\bibitem[{\citenamefont{Chen et~al.}(2005)\citenamefont{Chen, Zhang, Miao, Xu,
  Dong, Cao, Qiu, and Zhao}}]{6}
\bibinfo{author}{\bibfnamefont{W.~R.} \bibnamefont{Chen}},
  \bibinfo{author}{\bibfnamefont{F.~C.} \bibnamefont{Zhang}},
  \bibinfo{author}{\bibfnamefont{J.}~\bibnamefont{Miao}},
  \bibinfo{author}{\bibfnamefont{B.}~\bibnamefont{Xu}},
  \bibinfo{author}{\bibfnamefont{X.~L.} \bibnamefont{Dong}},
  \bibinfo{author}{\bibfnamefont{L.~X.} \bibnamefont{Cao}},
  \bibinfo{author}{\bibfnamefont{X.~G.} \bibnamefont{Qiu}}, \bibnamefont{and}
  \bibinfo{author}{\bibfnamefont{B.~R.} \bibnamefont{Zhao}},
  \bibinfo{journal}{Appl.Phys.Lett.} \textbf{\bibinfo{volume}{87}},
  \bibinfo{pages}{042508} (\bibinfo{year}{2005}).

\bibitem[{\citenamefont{B.Martinez et~al.}(1994)\citenamefont{B.Martinez,
  A.Labarta, R.Rodriguez-Sola, and X.Obradors}}]{7}
\bibinfo{author}{\bibnamefont{B.Martinez}},
  \bibinfo{author}{\bibnamefont{A.Labarta}},
  \bibinfo{author}{\bibnamefont{R.Rodriguez-Sola}}, \bibnamefont{and}
  \bibinfo{author}{\bibnamefont{X.Obradors}}, \bibinfo{journal}{Phys. Rev. B}
  \textbf{\bibinfo{volume}{50}}, \bibinfo{pages}{15779} (\bibinfo{year}{1994}).

\bibitem[{\citenamefont{A.P.Ramirez et~al.}(1990)\citenamefont{A.P.Ramirez,
  G.P.Espinosa, and A.S.Cooper}}]{8}
\bibinfo{author}{\bibnamefont{A.P.Ramirez}},
  \bibinfo{author}{\bibnamefont{G.P.Espinosa}}, \bibnamefont{and}
  \bibinfo{author}{\bibnamefont{A.S.Cooper}}, \bibinfo{journal}{Phys. Rev.
  Lett.} \textbf{\bibinfo{volume}{64}}, \bibinfo{pages}{2070}
  (\bibinfo{year}{1990}).

\bibitem[{\citenamefont{Hagemann et~al.}(2001)\citenamefont{Hagemann, Huang,
  Gao, Ramirez, and Cava}}]{9}
\bibinfo{author}{\bibfnamefont{I.~S.} \bibnamefont{Hagemann}},
  \bibinfo{author}{\bibfnamefont{Q.}~\bibnamefont{Huang}},
  \bibinfo{author}{\bibfnamefont{X.~P.~A.} \bibnamefont{Gao}},
  \bibinfo{author}{\bibfnamefont{A.~P.} \bibnamefont{Ramirez}},
  \bibnamefont{and} \bibinfo{author}{\bibfnamefont{R.~J.} \bibnamefont{Cava}}
  (\bibinfo{year}{2001}).

\bibitem[{\citenamefont{Chung et~al.}(2005)\citenamefont{Chung, Matsuda, Lee,
  Kakurai, Ueda, Sato, Takagi, Hong, and Park}}]{10}
\bibinfo{author}{\bibfnamefont{J.-H.} \bibnamefont{Chung}},
  \bibinfo{author}{\bibfnamefont{M.}~\bibnamefont{Matsuda}},
  \bibinfo{author}{\bibfnamefont{S.-H.} \bibnamefont{Lee}},
  \bibinfo{author}{\bibfnamefont{K.}~\bibnamefont{Kakurai}},
  \bibinfo{author}{\bibfnamefont{H.}~\bibnamefont{Ueda}},
  \bibinfo{author}{\bibfnamefont{T.~J.} \bibnamefont{Sato}},
  \bibinfo{author}{\bibfnamefont{H.}~\bibnamefont{Takagi}},
  \bibinfo{author}{\bibfnamefont{K.-P.} \bibnamefont{Hong}}, \bibnamefont{and}
  \bibinfo{author}{\bibfnamefont{S.}~\bibnamefont{Park}},
  \bibinfo{journal}{Phys. Rev. Lett.} \textbf{\bibinfo{volume}{95}},
  \bibinfo{pages}{247204} (\bibinfo{year}{2005}).

\bibitem[{\citenamefont{Rovers et~al.}(2002)\citenamefont{Rovers, Kyriakou,
  Dabkowska, Luke, Larkin, and Savici}}]{12}
\bibinfo{author}{\bibfnamefont{M.~T.} \bibnamefont{Rovers}},
  \bibinfo{author}{\bibfnamefont{P.~P.} \bibnamefont{Kyriakou}},
  \bibinfo{author}{\bibfnamefont{H.~A.} \bibnamefont{Dabkowska}},
  \bibinfo{author}{\bibfnamefont{G.~M.} \bibnamefont{Luke}},
  \bibinfo{author}{\bibfnamefont{M.~I.} \bibnamefont{Larkin}},
  \bibnamefont{and} \bibinfo{author}{\bibfnamefont{A.~T.}
  \bibnamefont{Savici}}, \bibinfo{journal}{Phys. Rev. B}
  \textbf{\bibinfo{volume}{66}}, \bibinfo{pages}{174434}
  (\bibinfo{year}{2002}).

\bibitem[{\citenamefont{M.Matsuda et~al.}(2007)\citenamefont{M.Matsuda,
  M.Takeda, M.Nakamura, Kakurai, Oosawa, Leli¨¨vre-Berna, Chung, Ueda, Takagi,
  and Lee}}]{11}
\bibinfo{author}{\bibnamefont{M.Matsuda}},
  \bibinfo{author}{\bibnamefont{M.Takeda}},
  \bibinfo{author}{\bibnamefont{M.Nakamura}},
  \bibinfo{author}{\bibfnamefont{K.}~\bibnamefont{Kakurai}},
  \bibinfo{author}{\bibfnamefont{A.}~\bibnamefont{Oosawa}},
  \bibinfo{author}{\bibfnamefont{E.}~\bibnamefont{Leli¨¨vre-Berna}},
  \bibinfo{author}{\bibfnamefont{J.-H.} \bibnamefont{Chung}},
  \bibinfo{author}{\bibfnamefont{H.}~\bibnamefont{Ueda}},
  \bibinfo{author}{\bibfnamefont{H.}~\bibnamefont{Takagi}}, \bibnamefont{and}
  \bibinfo{author}{\bibfnamefont{S.-H.} \bibnamefont{Lee}},
  \bibinfo{journal}{Phys. Rev. B} \textbf{\bibinfo{volume}{75}},
  \bibinfo{pages}{104415} (\bibinfo{year}{2007}).

\bibitem[{\citenamefont{L.Q.Yan et~al.}(2007)\citenamefont{L.Q.Yan, Jiang,
  X.D.Peng, L.H.He, and F.W.Wang}}]{13}
\bibinfo{author}{\bibnamefont{L.Q.Yan}},
  \bibinfo{author}{\bibfnamefont{Z.}~\bibnamefont{Jiang}},
  \bibinfo{author}{\bibnamefont{X.D.Peng}},
  \bibinfo{author}{\bibnamefont{L.H.He}}, \bibnamefont{and}
  \bibinfo{author}{\bibnamefont{F.W.Wang}}, \bibinfo{journal}{Powder Diffr.}
  \textbf{\bibinfo{volume}{22}}, \bibinfo{pages}{320} (\bibinfo{year}{2007}).

\bibitem[{\citenamefont{Dho et~al.}(2002)\citenamefont{Dho, Kim, and Hur}}]{14}
\bibinfo{author}{\bibfnamefont{J.}~\bibnamefont{Dho}},
  \bibinfo{author}{\bibfnamefont{W.~S.} \bibnamefont{Kim}}, \bibnamefont{and}
  \bibinfo{author}{\bibfnamefont{N.}~\bibnamefont{Hur}},
  \bibinfo{journal}{Phys. Rev. Lett.} \textbf{\bibinfo{volume}{89}},
  \bibinfo{pages}{027202} (\bibinfo{year}{2002}).

\bibitem[{\citenamefont{Prince}(1957)}]{15}
\bibinfo{author}{\bibfnamefont{E.}~\bibnamefont{Prince}},
  \bibinfo{journal}{Acta Crystallogr.} \textbf{\bibinfo{volume}{10}},
  \bibinfo{pages}{554} (\bibinfo{year}{1957}).

\bibitem[{\citenamefont{Jonason et~al.}(1996)\citenamefont{Jonason, Mattsson,
  and Nordblad}}]{16}
\bibinfo{author}{\bibfnamefont{K.}~\bibnamefont{Jonason}},
  \bibinfo{author}{\bibfnamefont{J.}~\bibnamefont{Mattsson}}, \bibnamefont{and}
  \bibinfo{author}{\bibfnamefont{P.}~\bibnamefont{Nordblad}},
  \bibinfo{journal}{Phys. Rev. B} \textbf{\bibinfo{volume}{53}},
  \bibinfo{pages}{6507} (\bibinfo{year}{1996}).

\bibitem[{\citenamefont{Sullow et~al.}(1997)\citenamefont{Sullow, Nieuwenhuys,
  A.A.Menovsky, J.A.Mydosh, S.A.M.Mentink, T.E.Mason, and W.J.L.Buyers}}]{17}
\bibinfo{author}{\bibfnamefont{S.}~\bibnamefont{Sullow}},
  \bibinfo{author}{\bibfnamefont{G.~J.} \bibnamefont{Nieuwenhuys}},
  \bibinfo{author}{\bibnamefont{A.A.Menovsky}},
  \bibinfo{author}{\bibnamefont{J.A.Mydosh}},
  \bibinfo{author}{\bibnamefont{S.A.M.Mentink}},
  \bibinfo{author}{\bibnamefont{T.E.Mason}}, \bibnamefont{and}
  \bibinfo{author}{\bibnamefont{W.J.L.Buyers}}, \bibinfo{journal}{Phys. Rev.
  Lett.} \textbf{\bibinfo{volume}{78}}, \bibinfo{pages}{354}
  (\bibinfo{year}{1997}).

\bibitem[{\citenamefont{K.Gunnarsson et~al.}(1988)\citenamefont{K.Gunnarsson,
  P.Svedlindh, P.Nordblad, L.Lundgren, H.Aruga, and A.Ito}}]{18}
\bibinfo{author}{\bibnamefont{K.Gunnarsson}},
  \bibinfo{author}{\bibnamefont{P.Svedlindh}},
  \bibinfo{author}{\bibnamefont{P.Nordblad}},
  \bibinfo{author}{\bibnamefont{L.Lundgren}},
  \bibinfo{author}{\bibnamefont{H.Aruga}}, \bibnamefont{and}
  \bibinfo{author}{\bibnamefont{A.Ito}}, \bibinfo{journal}{Phys. Rev. Lett.}
  \textbf{\bibinfo{volume}{61}}, \bibinfo{pages}{754} (\bibinfo{year}{1988}).

\bibitem[{\citenamefont{Fisher and H.Sompolinsky}(1985)}]{19}
\bibinfo{author}{\bibfnamefont{D.}~\bibnamefont{Fisher}} \bibnamefont{and}
  \bibinfo{author}{\bibnamefont{H.Sompolinsky}}, \bibinfo{journal}{Phys. Rev.
  Lett.} \textbf{\bibinfo{volume}{54}}, \bibinfo{pages}{1063}
  (\bibinfo{year}{1985}).

\bibitem[{\citenamefont{D.N.H.Nam et~al.}(1999)\citenamefont{D.N.H.Nam,
  K.Jonason, P.Nordblad, N.V.Khiem, and N.X.Phuc}}]{20}
\bibinfo{author}{\bibnamefont{D.N.H.Nam}},
  \bibinfo{author}{\bibnamefont{K.Jonason}},
  \bibinfo{author}{\bibnamefont{P.Nordblad}},
  \bibinfo{author}{\bibnamefont{N.V.Khiem}}, \bibnamefont{and}
  \bibinfo{author}{\bibnamefont{N.X.Phuc}}, \bibinfo{journal}{Phys. Rev. B}
  \textbf{\bibinfo{volume}{59}}, \bibinfo{pages}{4189} (\bibinfo{year}{1999}).

\bibitem[{\citenamefont{Kunkel et~al.}(1988)\citenamefont{Kunkel, Roshko, and
  G.Williams}}]{21}
\bibinfo{author}{\bibfnamefont{H.}~\bibnamefont{Kunkel}},
  \bibinfo{author}{\bibfnamefont{R.~M.} \bibnamefont{Roshko}},
  \bibnamefont{and} \bibinfo{author}{\bibnamefont{G.Williams}},
  \bibinfo{journal}{Phys. Rev. B} \textbf{\bibinfo{volume}{37}},
  \bibinfo{pages}{5880} (\bibinfo{year}{1988}).

\bibitem[{\citenamefont{Arrott}(1957)}]{22}
\bibinfo{author}{\bibfnamefont{A.}~\bibnamefont{Arrott}},
  \bibinfo{journal}{Phys. Rev.} \textbf{\bibinfo{volume}{108}},
  \bibinfo{pages}{1394} (\bibinfo{year}{1957}).

\bibitem[{\citenamefont{Yeung et~al.}(1986)\citenamefont{Yeung, Roshko, and
  Williams}}]{23}
\bibinfo{author}{\bibfnamefont{I.}~\bibnamefont{Yeung}},
  \bibinfo{author}{\bibfnamefont{R.~M.} \bibnamefont{Roshko}},
  \bibnamefont{and} \bibinfo{author}{\bibfnamefont{G.}~\bibnamefont{Williams}},
  \bibinfo{journal}{Phys. Rev. B} \textbf{\bibinfo{volume}{34}},
  \bibinfo{pages}{3456} (\bibinfo{year}{1986}).

\bibitem[{24()}]{24}
\emph{\bibinfo{title}{No attempt was made to use the neutron scattering
  experiments to investigate the magnetic phase separation in our case because
  of the high neutron-capture cross section of \protect{C}$d^{113}$ isotope
  (about 2000 bars)}}.

\bibitem[{\citenamefont{Urano et~al.}(2000)\citenamefont{Urano, Nohara, Kondo,
  Sakai, Takagi, Shiraki, and T.Okubo}}]{25}
\bibinfo{author}{\bibfnamefont{C.}~\bibnamefont{Urano}},
  \bibinfo{author}{\bibfnamefont{M.}~\bibnamefont{Nohara}},
  \bibinfo{author}{\bibfnamefont{S.}~\bibnamefont{Kondo}},
  \bibinfo{author}{\bibfnamefont{F.}~\bibnamefont{Sakai}},
  \bibinfo{author}{\bibfnamefont{H.}~\bibnamefont{Takagi}},
  \bibinfo{author}{\bibfnamefont{T.}~\bibnamefont{Shiraki}}, \bibnamefont{and}
  \bibinfo{author}{\bibnamefont{T.Okubo}}, \bibinfo{journal}{Phys. Rev. Lett.}
  \textbf{\bibinfo{volume}{85}}, \bibinfo{pages}{1052} (\bibinfo{year}{2000}).

\end{thebibliography}
\end{document}